\documentstyle[epsfig]{mn2e}
\begin{document}
\input BoxedEPS.tex
\newcommand{\aip}{{\small ${\cal AIPS}$}}
\newcommand{\gtsim}{\mbox{{\raisebox{-0.4ex}{$\stackrel{>}{{\scriptstyle\sim}}
$}}}}
\newcommand{\ltsim}{\mbox{{\raisebox{-0.4ex}{$\stackrel{<}{{\scriptstyle\sim}}
$}}}}
\newcommand{\s}{$\stackrel{\rm s}{.}$}
\newcommand{\h}{$^{\rm h}$}
\newcommand{\m}{$^{\rm m}$}
\newcommand{\pp}{$\stackrel{\prime\prime}{.}$}
\newcommand{\de}{$^{\circ}$}
\newcommand{\p}{$^{\prime}$}
\newcommand{\arc}{$^{\prime\prime}$}
\newcommand{\marc}{^{\prime\prime}}
\newcommand{\rs}{{\em $r_s$}}
\newcommand{\DPM}{{\em DPM}}
\newcommand{\alf}{{\displaystyle\biggl({\nu_{\rm h} \over \nu_{\rm l}}\biggr)^{\alpha}} }

\newcommand{\figstart}[1]
    { \begin{figure}[htb]
      \begin{picture}(0,#1) }
\newcommand{\figend}[4]
    { \end{picture}
      \special{#1}
      \caption[#2]{#3}
      \label{#4}
      \end{figure} }
\newcommand{\fig}[5]
    { \figstart{#1}
      \figend{#2}{#3}{#4}{#5} }
\newcommand{\bHS}{\beta_{\mbox{\scriptsize HS}}}
\newcommand{\bBF}{\beta_{\mbox{\scriptsize BF}}}
\newcommand{\nT}{\nu_{\mbox{\scriptsize T}}}
\newcommand{\et}{E_{\mbox{\scriptsize T}}}
\newcommand{\nTn}{\nu_{\mbox{\scriptsize Tn}}}
\newcommand{\nTf}{\nu_{\mbox{\scriptsize Tf}}}
\newcommand{\tn}{\tau_{x\mbox{\scriptsize n}}}
\newcommand{\tf}{\tau_{x\mbox{\scriptsize f}}}
\newcommand{\xn}{x_{\mbox{\scriptsize n}}}
\newcommand{\xf}{x_{\mbox{\scriptsize f}}}
\newcommand{\yn}{y_{\mbox{\scriptsize n}}}
\newcommand{\yf}{y_{\mbox{\scriptsize f}}}
\newcommand{\lln}{l_{\mbox{\scriptsize n}}}
\newcommand{\llf}{l_{\mbox{\scriptsize f}}}
\newcommand{\Dn}{f(\Delta_{\mbox{\scriptsize n}})}
\newcommand{\Df}{f(\Delta_{\mbox{\scriptsize f}})}
\newcommand{\B}{\mbox{$B$}}
\newcommand{\Bo}{\mbox{$B$}_{0}}

\SetEPSFDirectory{/scratch/sbgs/figures/hst/}
\SetRokickiEPSFSpecial
\HideDisplacementBoxes

\title[A new model for infrared and submillimetre counts]{A new model for infrared and submillimetre counts}
\author[Rowan-Robinson M.]{Michael Rowan-Robinson\\
Astrophysics Group, Blackett Laboratory, Imperial College of Science 
Technology and Medicine, Prince Consort Road,\\ 
London SW7 2AZ}
\maketitle
\begin{abstract}
A new model for source counts from 8-1100 $\mu$m is presented, which agrees well with source-count
data and the observed background spectrum.  The model is similar to that of Rowan-Robinson (2001), but
with different evolution for each of the four assumed infrared template types.   The evolution is modified in two
ways; (i) the exponential factor is modified so that it tends to a constant value at late times, (ii) the power-law
factor is modified so that it tends to zero at redshift $z_f$, rather than 0 as assumed by Rowan-Robinson (2001).
I find strong evidence from the 850 and 1100 $\mu$m counts, and from the infrared background, that $z_f$ = 4-5, with
some preference for a value at the low end of the range, implying
that star-forming galaxies at z $>$ 5 are not significant infrared emitters, presumably due to a low opacity in
dust at these early epochs.

The model involves zero or even negative evolution for starbursts and AGN at low redshifts ($<$0.2), suggesting
that the era of major mergers and strong galaxy-galaxy interactions is over.

\end{abstract}

%\large

\section{Introduction}

Source-counts at infrared and submillimetre wavelengths, combined with the spectrum of the 
infrared background, give us important constraints on the star-formation history of the universe.
First indications of strong evolution in the properties of star-forming galaxies came from IRAS 
60 $\mu$m counts (Hacking and Houck 1987, Lonsdale et al 1990, Rowan-Robinson et al 1991,
Franceschini et al 1991, 1994, Pearson and Rowan-Robinson 1996)

ISO gave us deep counts at 15 $\mu$m providing strong evidence for evolution
(Oliver et al 1997,  Rowan-Robinson et al 1997,  Guiderdoni et al 1998, Aussel et al 1999, 
Elbaz et al 1999, 2002, Serjeant et al 2000, Gruppioni et al 2002, Lagache et al 2003) and useful 
counts at 90 and 175 $\mu$m
(Kawara et al 1998, Dole et al 2001, Efstathiou et al 2000a, Heraudeau et al 2004).

850 $\mu$m counts with SCUBA on JCMT (Smail et al 1997, Hughes et al 1998, 
Eales et al 1999, 2000, Barger et al 1999, Blain et al 1999, Fox et al 2002, Scott et al 2002,
Smail et al 2002, Cowie et al 2002, Webb et al 2003, Borys et al 2004, Coppin et al 2006) have given
 important insight into the 
role of cool dust and also strong constraints on the high-redshift evolution of hyperluminous infrared galaxies. 
 Evidence for luminosity evolution of submillimetre galaxies was given by Ivison et al (2002).
1200 $\mu$m counts with MAMBO have been reported by Greve et al (2004) and modelled in terms
of a strongly evolving population.  
Recently the AZTEC collaboration have reported differential counts at 1100 $\mu$m (Perera et al 2008,
Austermann et al 2008), which provide even stronger constraints on high redshift evolution.

The detection of the infrared background with COBE (Puget et al 1996, Fixsen et al 1998, Hauser et al 1998, 
Lagache et al 1999) provided an important constraint on evolutionary models
(Guiderdoni et al 1997, 1998, Franceschini et al 1998, 2001, Dwek et al 1998, Blain et al 1999, Gispert et al 2000,
Dole et al 2001, Rowan-Robinson 2001,  Chary and Elbaz 2001, Elbaz et al 2002,  King and RR 2003,
Balland et al 2003, Xu et al 2003, Lagache et al 2003).  Dole et al (2006) used stacking analysis on deep
{\it Spitzer} counts at 24, 90 and 160 $\mu$m to estimate the integrated background radiation from sources
at these wavelengths.  With {\it Spitzer} we also have deep counts at 8, 24, 70, 160 $\mu$m (Fazio et al 2004, 
Chary et al 2004, Marleau et al 2004, Papovich et al 2004, Dole et al 2004, Le Floch et al 2005, 
Frayer et al 2006, Shupe et al 2008), with 24 $\mu$m providing 
an especially complete picture of the contribution of individual sources to the  infrared 
background.  While some pre-{\it Spitzer} models were quite successful at 70 and 160 $\mu$m,
none captured full details of the 24 $\mu$m counts.  Lagache et al (2004) provided early revised models
in the light of the Spitzer data.  

Rowan-Robinson (2001) modeled infrared source-counts and background in terms of four
types of infrared galaxy: quiescent galaxies in which the infrared radiation (infrared 'cirrus')
is emission from interstellar dust illuminated by the general stellar radiation field, starbursts
for which M82 is the prototype, more extreme, higher optical depth starbursts with Arp 220 as 
prototype, and AGN dust tori.  The same evolution history, intended to reflect the global star-formation
history, was used for each galaxy type.  This model was consistent with counts, luminosity 
functions and colour-luminosity relations from IRAS, counts from 
ISO and  available integral 850 $\mu$m counts.  However the advent of deep source-count data from
{\it Spitzer} showed that this model, along with other pre-{\it Spitzer} models, failed, especially 
at 24 $\mu$m.  

In this paper I show how the Rowan-Robinson (2001) model has to be modified to achieve consistency
with {\it Spitzer} and other modern data and give predictions for the {\it Herschel} and {\it Planck} wave-bands.
Some preliminary results were given in Shupe et al (2008).

A cosmological model with $\Lambda$ = 0.7, $h_0$=0.72 has been used throughout.

\section{Methodology}

The philosophy is similar to that of Rowan-Robinson (2001), to find a simple analytic form for the 
evolutionary function, without discontinuities and involving the minimum number of parameters.
I retain the same four basic infrared templates: (1) quiescent galaxies, radiating in the infrared through 
emission by interstellar dust of absorbed starlight (' infrared cirrus'), (2) starbursts, with prototype 
M82, (3) extreme starbursts with much higher dust optical depth, with prototype Arp 220, (4) AGN
dust tori.  These templates, which have been derived through radiative transfer calculations
(Efstathiou and Rowan-Robinson 1995, 2003, Rowan-Robinson 1995, Efstathiou et al 2000b), have proved 
extremely effective in modeling the spectral
energy distributions (SEDs) of infrared galaxies in the ISO-ELAIS (Rowan-Robinson et al 2004),
 {\it Spitzer}-SWIRE (Rowan-Robinson et al 2005) and SHADES (Clements et al 2008) surveys .   
The luminosity functions for each 
component are derived as in Rowan-Robinson (2001), from the 60 $\mu$m luminosity function
via a mixture table which is a function of 60 $\mu$m luminosity (Figure 1).  These
luminosity functions are essentially unchanged from Rowan-Robinson (2001), except that the
Arp 220 component has been assumed to be more significant in the luminosity range
$L_{60} = 10^9-10^{10} L_{\odot}$, in order to improve the fit to the counts at 850 $\mu$m
(see below).

\begin{figure*}
\epsfig{file=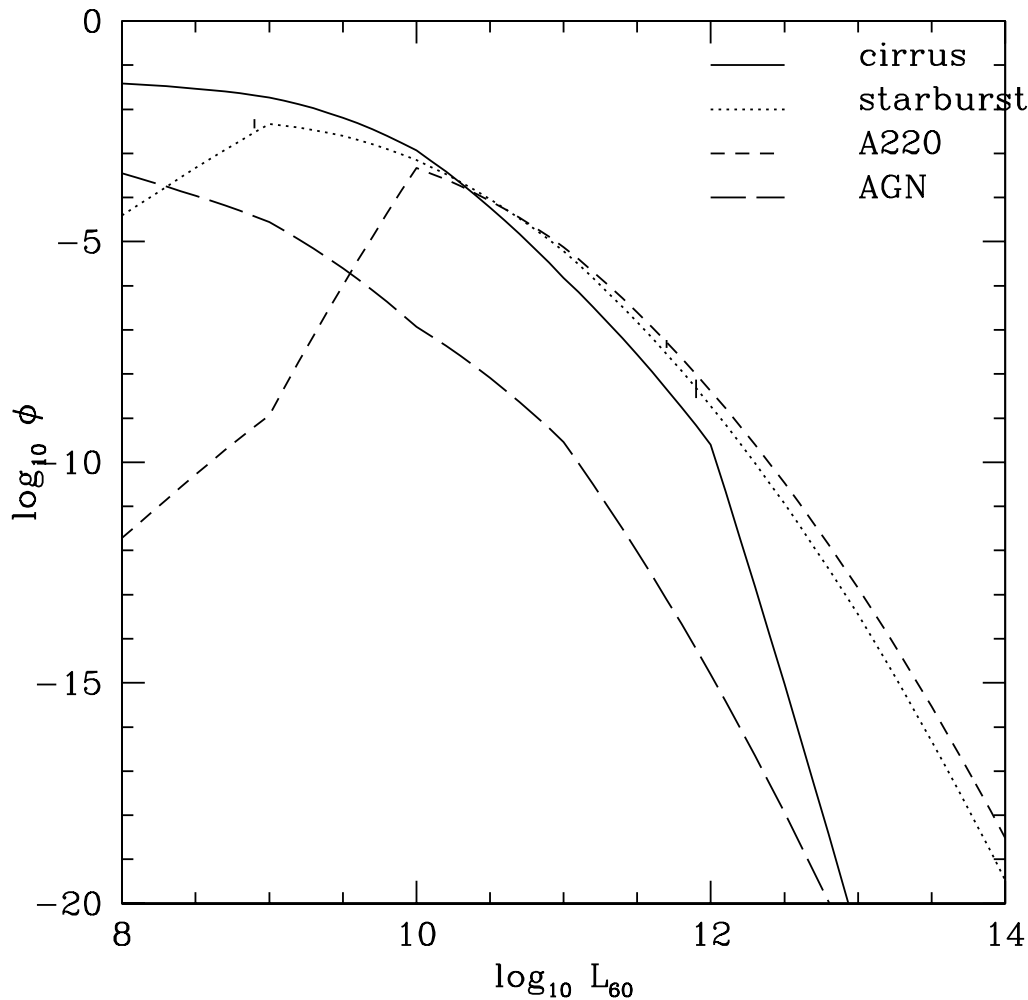,angle=0,width=14cm}
\caption{Contributions to 60 $\mu$m luminosity function of cirrus (solid), M82 (dotted), A220 (dashed), 
and AGN dust tori (long-dashed) infrared components. 
}
\end{figure*}

\begin{table}
\caption{Evolutionary parameters for each component}
\begin{tabular}{llll}
component & P& Q& $a_0$\\
&&&\\
cirrus & 4.0 & 10.0& 0.6\\
M82 & 2.7 & 18.0 & 0.994\\
A220 & 2.7 & 13.6 & 0.92\\
AGN & 2.7 & 13.0 & 0.94\\
\end{tabular}
\end{table}

The main new feature of the current work is that I now allow each component its own evolution 
function, since a single evolution function for all components is not capable of reproducing the
latest counts, especially at 24 $\mu$m.

In Rowan-Robinson (2001) I assumed a pure luminosity evolution, with $L(z) = \phi(z) L(0)$, and

$\phi(z) = exp { Q (t_0-t)/t_0 } (t/t_0)^P$		(1)

where $t_0$ is the current epoch, the exponential factor is essentially the Bruzual and Charlot  
(1993) star-formation history
and the power law factor ensures that L(z) tends to zero as t tends to 0.  Thus $t_0/Q$ has the meaning
of the exponential time-scale for star-formation.  The peak star-formation rate in this model
occurs at $t/t_0 = P/Q$.

However the 24 $\mu$m counts show an extended Euclidean behaviour at bright fluxes and 
I have therefore allowed for the possibility that the evolution function tends asymptotically to a
constant value at late times.  I also allow the epoch where the evolution function tends to zero
at early times to be $t_f = t(z_f)$ rather than zero, where $z_f$ is to be determined from the
fits to the counts.  This option was found to be important at submillimetre wavelengths.  Thus
I assume

$\phi(z) = [ a_0 + (1-a_0) exp  Q (1-y )]  (y^P -(y_f)^P)$                 (2)

where $y = (t/t_0)$, $y_f = t(z_f)/t_0$.

The strategy for determining the parameters P, Q, $a_0$  for each component was as follows.
First we allowed all parameters to vary while seeking a best-fit solution to the 24 $\mu$m
counts.  This fixed the parameters for the cirrus and M82 components.  Keeping these fixed
we set P for the AGN dust tori to have the same value as the M82 starburst component and
then optimized Q and $a_0$ for the dust tori to fit the 24 $\mu$m counts, while requiring that
the proportion of dust tori dominated sources at S(24) = 1 mJy be no less than 20 $\%$,
based on analysis of the SWIRE survey (Rowan-Robinson et al 2005).  These parameter 
choices essentially then
determined, without further adjustment, the counts at 8, 15, 70 and 160 $\mu$m and
the goodness of fit at these wavelengths is a test of the broad physical validity of
the assumed template and luminosity function.

The predicted counts at 850 (and 1100) $\mu$m depend sensitively on the evolutionary 
parameters for Arp 220 starburst component, but these parameters are essentially undetermined
at 24 $\mu$m.  It became clear that the proposal of Efstathiou and Rowan-Robinson (2003)
that cirrus galaxies could make a dominant contribution at S(850) = 8 mJy, was not consistent
with the count model approach adopted here, although they do become dominant at S(850) 
$<$ 1 mJy in these count models.  The evolutionary parameters for the A220 component
were adjusted to give an equal contribution of A220 and M82 starbursts at S(850) = 8 mJy,
since SED modelling of SHADES sources (Clements et al 2008) suggests both templates are important at
these fluxes .  This also required an increase in the luminosity function for A220 sources 
at lower luminosities.  We are not able to reduce the contribution of M82 starbursts in the submm
without destroying the fit to the counts at 8-160 $\mu$m.  Evidence that submillimetre sources
do not constitute a monolithic, Arp-220-like, population has been given by Menendez-Delmestre
et al (2007), Ibar et al (2008 and Pope et al (2008).

Adopted values of P, Q, $a_0$ for each of the four components are given in Table 1.
Figure 2 shows the adopted evolution functions for the four infrared components, 
with values of $z_f$ of 4, 5 and 10 illustrated.  The model suggests zero or even negative
evolution for some components at low redshifts and it would be interesting to test this
with large area far infrared surveys.  The physical interpretation would be that the era of
major mergers and strong interactions is over, with the remaining star-formation at the present 
epoch being relatively quiescent or driven by weaker interactions.

I should emphasize that the assumed evolution is applied to all infrared luminosities, and
I find no need to postulate very strong evolution only of luminous galaxies as has been 
proposed in many source-count models in the literature.

The assumption of a pure luminosity evolution is perhaps questionable, since we know that
galaxies grow by mergers, so that we expect some positive density evolution with redshift 
for more massive galaxies and negative evolution for lower masses.  However the evolution
we are trying to describe in eqn (2) refers only to star-forming, dusty galaxies.  If mergers involve
fragments which are not star-forming or have little dust, they will not contribute to density evolution
at these wavelengths.
It would be desirable in future work to relate evolutionary models to semi-analytical models 
for the merger of dark matter halos and the subsequent 
evolution of star-formation and dust, but this is beyond the scope of the present paper.

\begin{figure*}
\epsfig{file=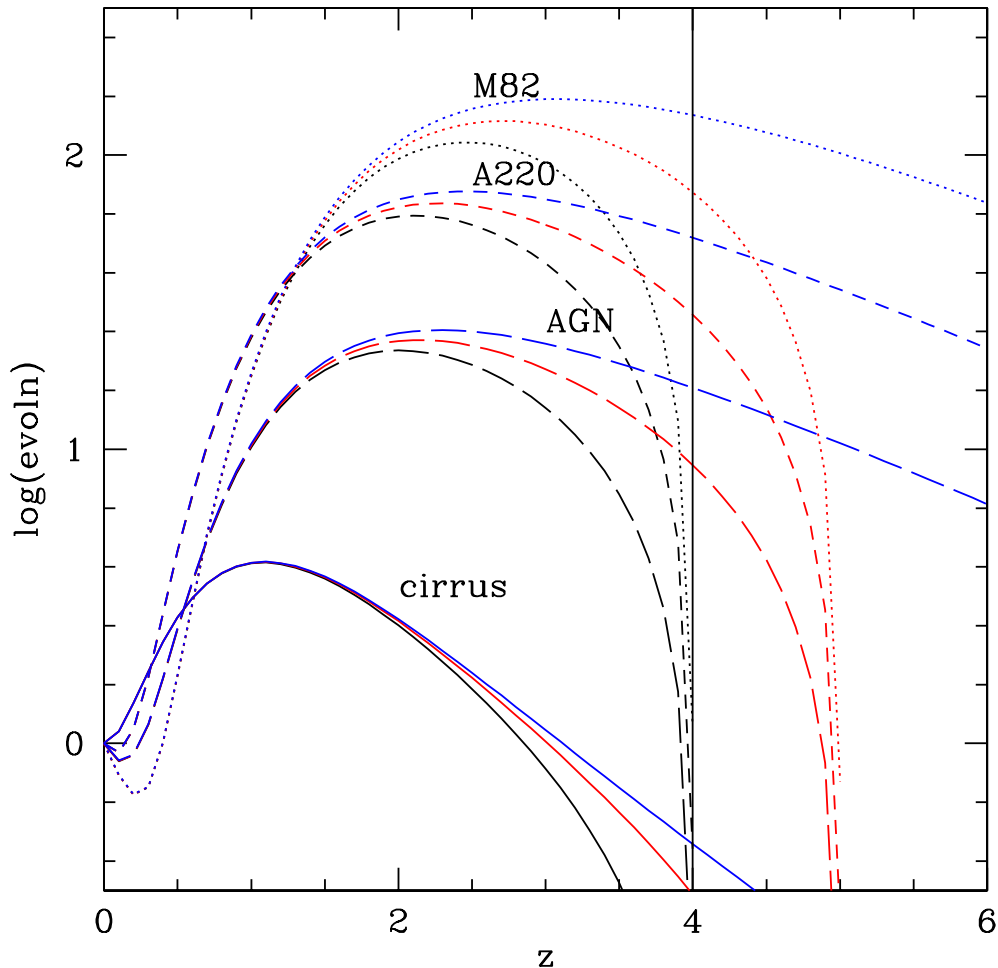,angle=0,width=14cm}
\caption{Evolution rate for cirrus (solid), M82 (dotted), A220 (dashed), and AGN dust tori
(long-dashed) infrared components. Black curves: adopted maximum redshift of 4; red curves: 
maximum redshift 5; blue curves: maximum redshift 10.
}
\end{figure*}

\section{Fits to counts}

Here I discuss the fits to counts at 8-1100 $\mu$m.  The most challenging wavelength is 24 $\mu$m,
with a relatively Euclidean behaviour at bright fluxes, a very sharp steepening of the counts 
between 3 mJy and 300 $\mu$Jy, and a strong flattening of the source-count slope fainter
than 200 $\mu$Jy. Figure 3 shows my fit to the {\it Spitzer} 24 $\mu$m differential (Euclidean-normalized) 
counts, indicating the contribution of each of the four infrared spectral types.  Both cirrus and
M82 starbursts are important at bright fluxes, but the strong steepening of the counts is
attributed solely to M82 starbursts.  AGN dust tori make their strongest contribution at 1-10 mJy.
Arp220 starburst make only a small contribution at all fluxes.

The 15 $\mu$m predicted differential counts (Fig 4R) fitted to ISO data show similar characteristics 
to the 24 $\mu$m counts
but the interpretation is quite different with cirrus galaxies dominant at all fluxes.  This reflects the strong 
difference between the spectral energy distributions (SEDs) of cirrus and M82 templates in the mid 
infrared.  The 8 $\mu$m predicted differential counts are compared to Spitzer data (Fazio et al 2006) in Fig 4L.
These appear quite different from the 15 and 24 $\mu$m counts, because they are far more strongly
dominated by quiescent galaxies.

\begin{figure*}
\epsfig{file=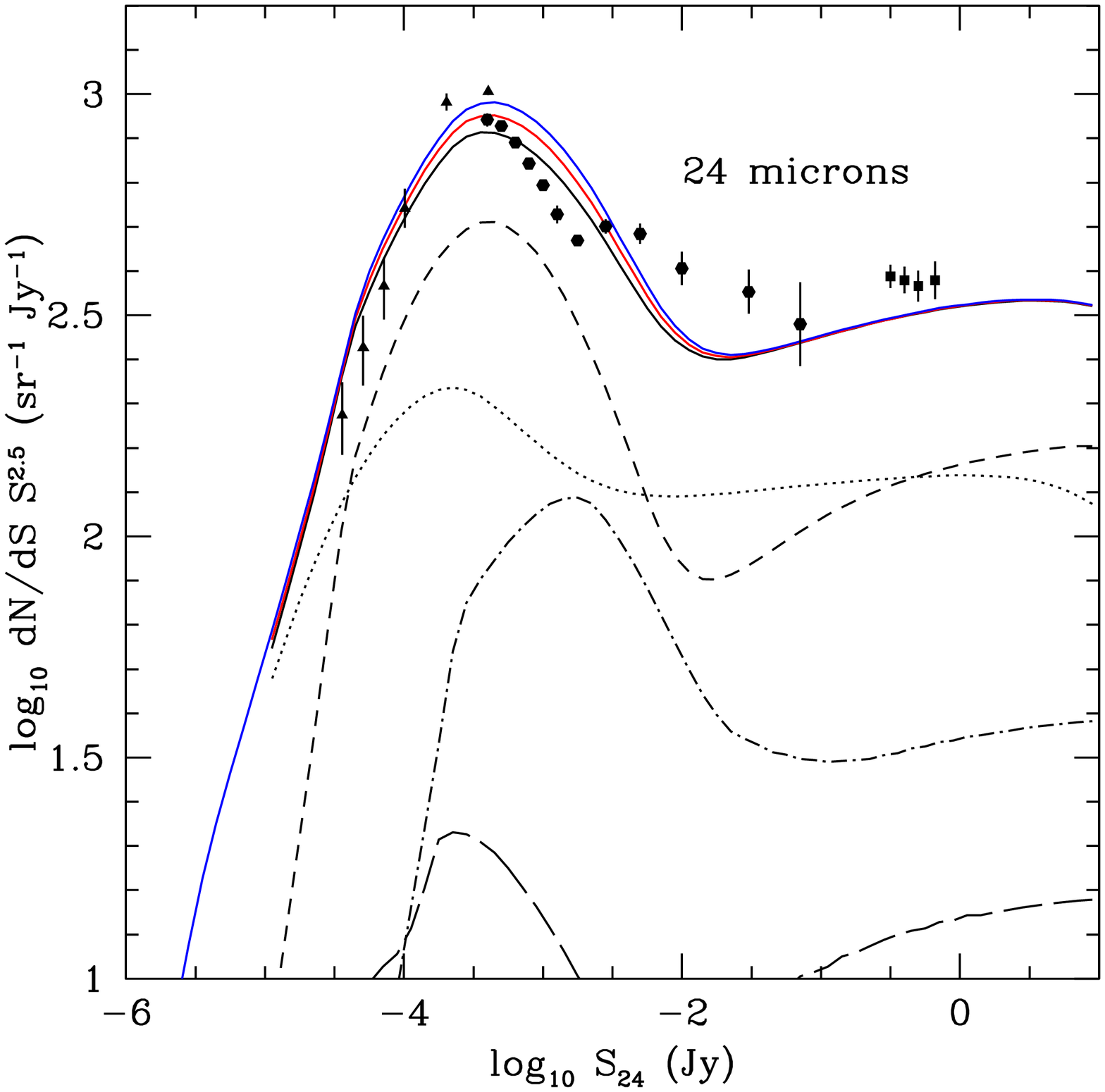,angle=0,width=14cm}
\caption{Euclidean normalised differential counts  24  $\mu$m.
Solid curve: total counts (black: $z_f$ = 4, red(upper): $z_f$ = 5) ; dotted curve: cirrus; short-dashed curve: M82;
long-dashed curve: A220; dash-dotted curve: and AGN dust tori.
Black curves: maximum redshift 4; red curve: maximum redshift 5, blue curve: maximum redshift 10.
Filled circles: SWIRE data from Shupe et al (2008), triangles: data from Papovich et al (2004). 
}
\end{figure*}

\begin{figure*}
\epsfig{file=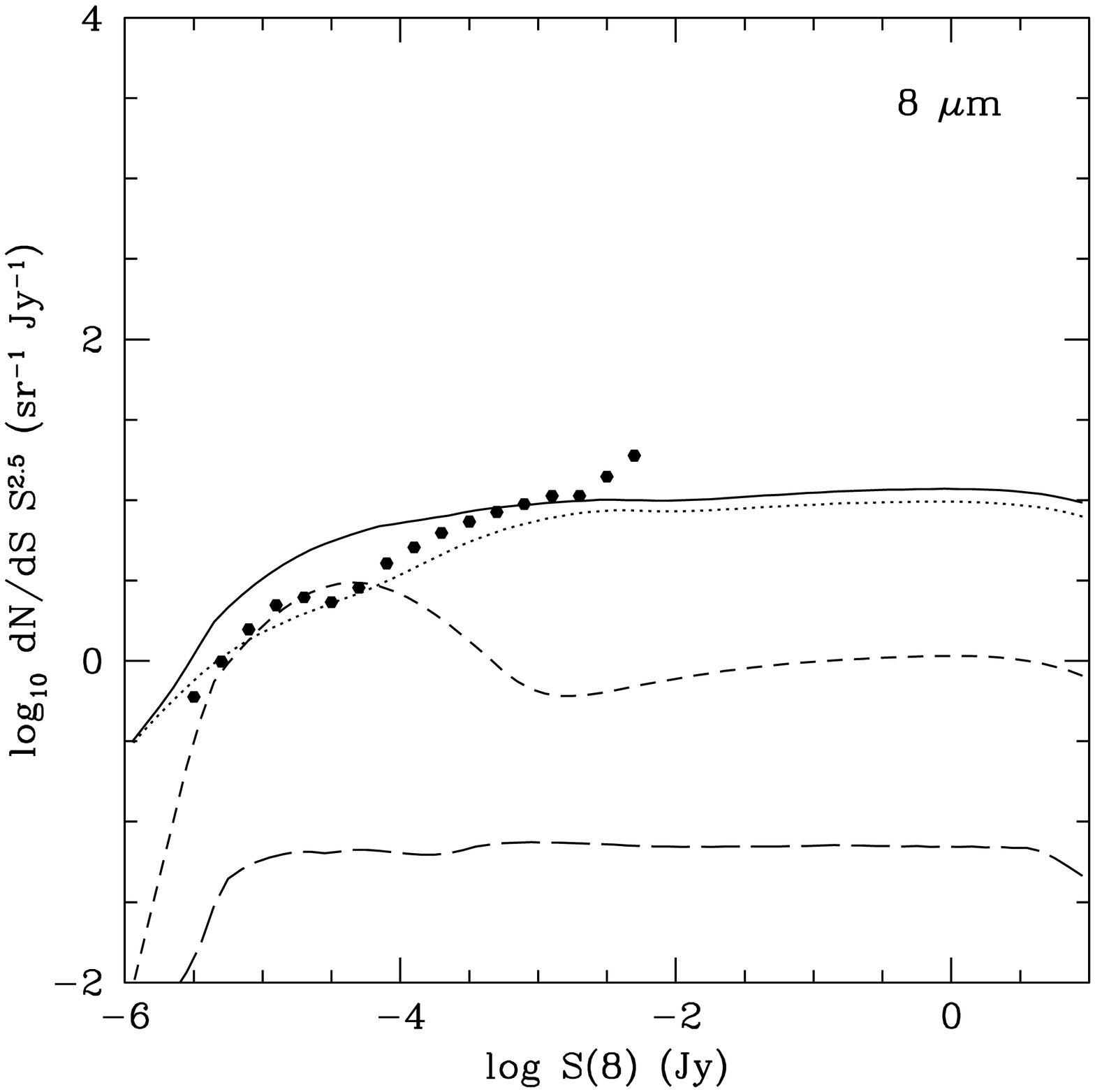,angle=0,width=7cm}
\epsfig{file=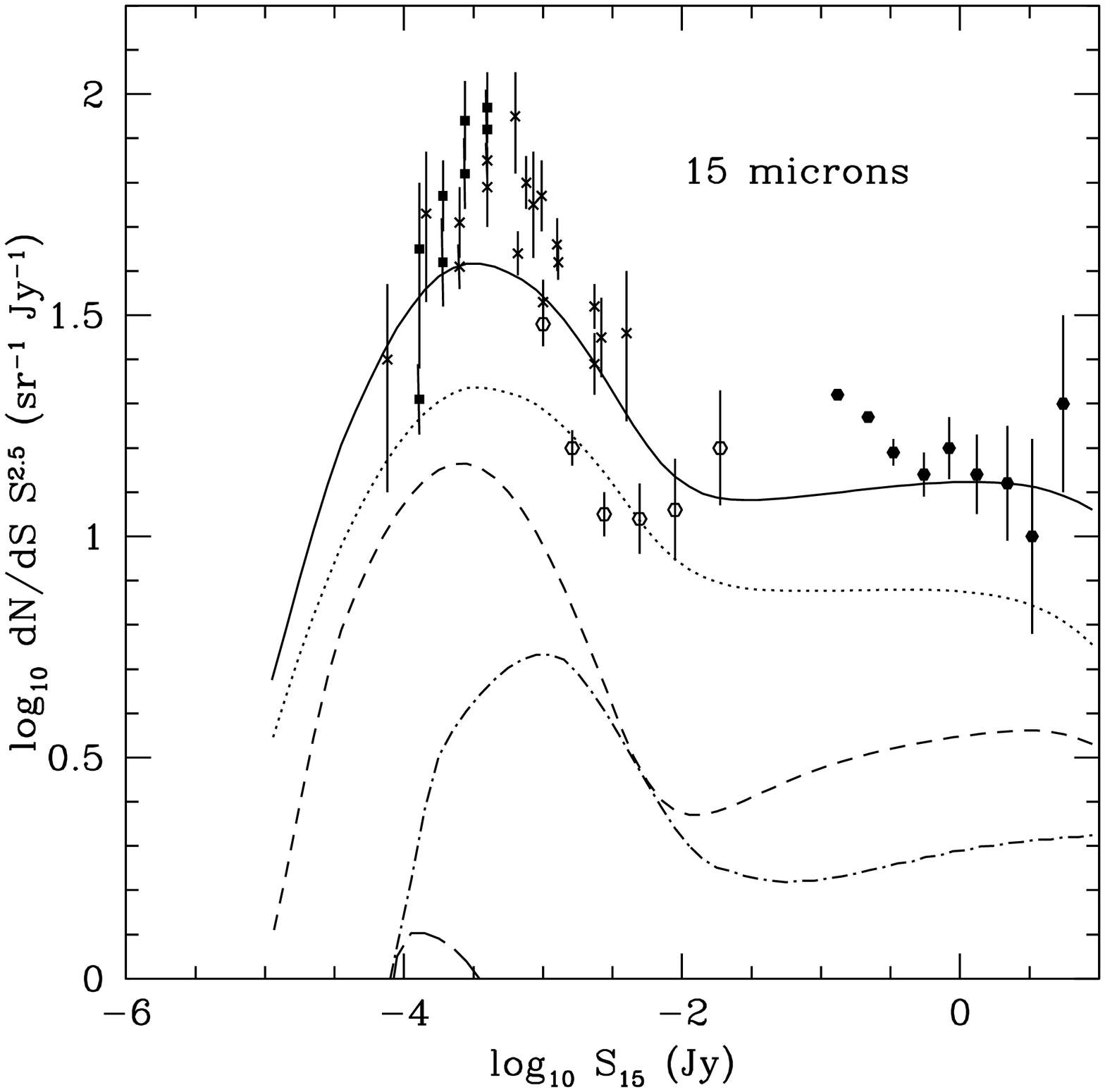,angle=0,width=7cm}
\caption{Euclidean normalised differential counts at 8 (LH) (data from Fazio et al 2004) and 15 (RH) $\mu$m
(ISO data from Elbaz et al 1999 (open triangles), Aussel et al 1999 (filled triangles), Gruppioni et al 2002
(open circles); interpolated IRAS data (filled circles) from Verma (private communication).
Solid curve: total counts; dotted curve: cirrus; short-dashed curve: M82;
long-dashed curve: A220; dash-dotted curve: AGN dust tori.
}
\end{figure*}

Figure 5 shows the fit to {\it Spitzer} counts at 70 and 160 $\mu$m differential counts.  Here we see
dominance by cirrus galaxies at bright fluxes and M82 starbursts at faint fluxes

\begin{figure*}
\epsfig{file=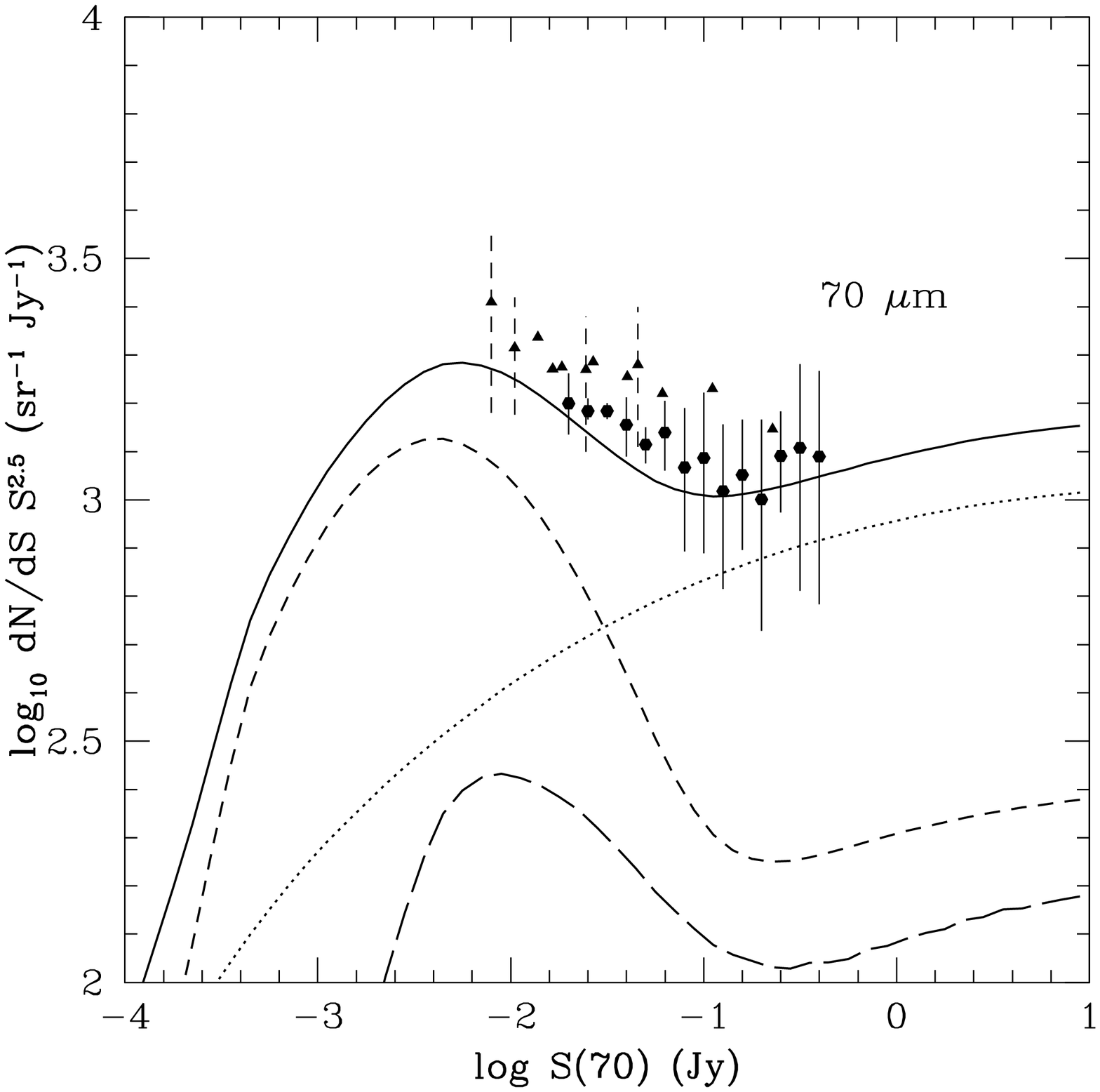,angle=0,width=7cm}
\epsfig{file=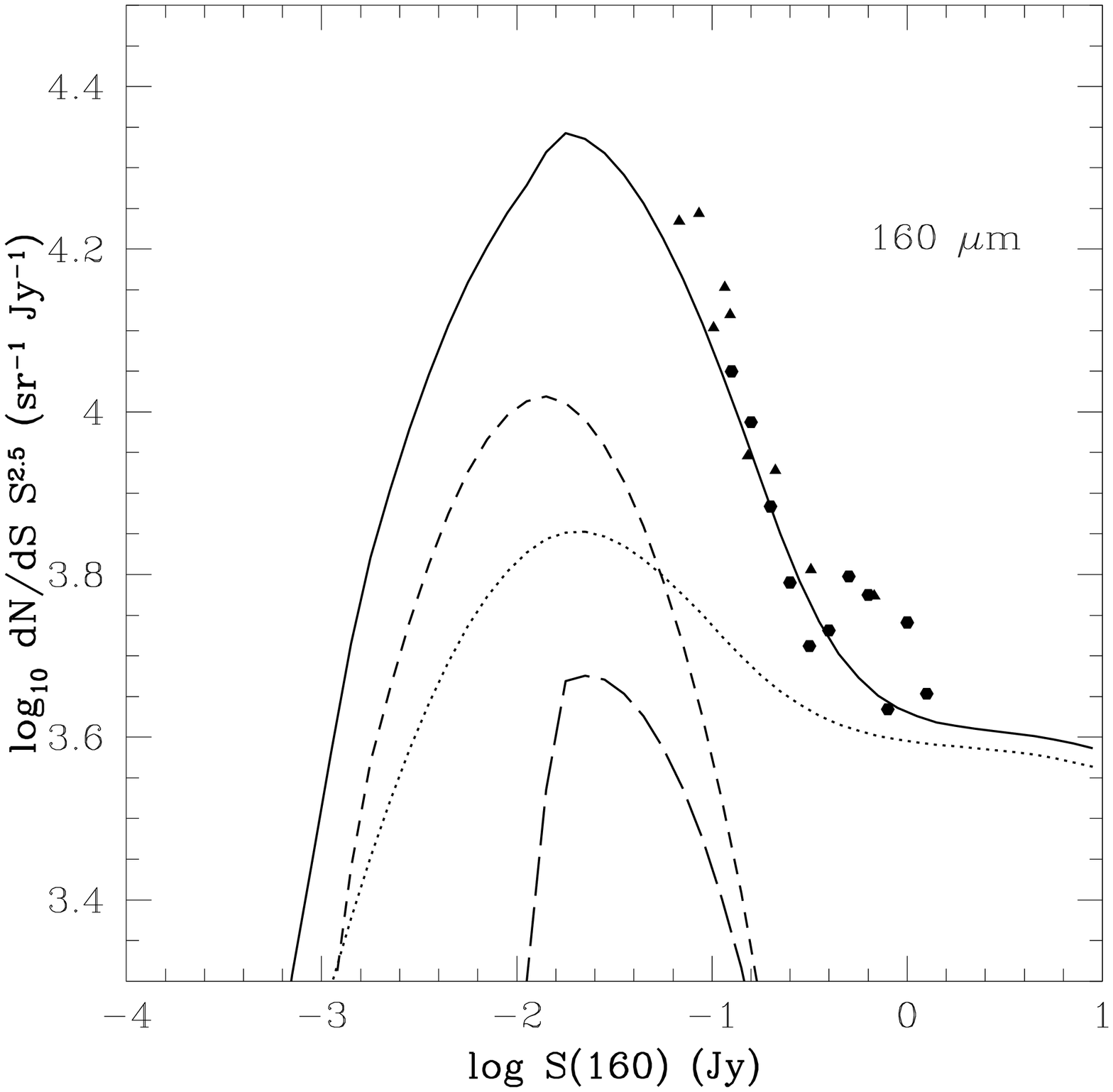,angle=0,width=7cm}
\caption{Euclidean normalised differential counts at 70 (LH) and 160 (RH) $\mu$m (filled circles: 
SWIRE Afonso-Luis et al (2008
in preparation), triangles: FLS Frayer et al 2006) .
Solid curve: total counts; dotted curve: cirrus; short-dashed curve: M82;
long-dashed curve: A220; dash-dotted curve: AGN dust tori.
Black curves: maximum redshift 4; red curve: maximum redshift 5.
}
\end{figure*}

Figure 6L shows the fit to SCUBA differential counts at 850 $\mu$m (Coppin et al 2006).  Cirrus 
galaxies are predicted to dominate at the very highest fluxes, though we have no source-count 
information at these fluxes yet.  The {\it Planck} mission should remedy this.  At the fainter fluxes sampled
by SCUBA surveys I predict that Arp220 starbursts should dominate above 10 mJy and
M82 starbursts below this.  This is consistent with the SED modelling of SHADES sources
by Clements et al (2008), which shows both templates being needed to fit the multiwavelength
SEDs.  There is a strong dependence on the redshift at which infrared starbursts are assumed to
switch on, $z_f$, with values in the range 4 to 5 favoured by 850 $\mu$m counts.   

Figure 6R shows the fit to the GOODS  1100 $\mu$m differential counts (Perera et al 2008).
%AZTEC/SHADES 1100 $\mu$m differential counts (Austermann et al 2008).  
These are similar
to the 850 $\mu$m counts in interpretation, with $z_f$ = 4-5 again being favoured.  This is at
first sight surprising since there is clear evidence now of galaxies at redshifts 6-7.  However it is
consistent with the 850 $\mu$m spectroscopic redshift  compilation of Chapman et al (2003, 2005), which 
shows no redshifts $>$ 4.  That work was subject to the constraint that the 850 $\mu$m sources
be detected at radio wavelengths and this could affect the ability to detect high redshift objects, but
Chapman et al (2005) argue that this is not a major selection effect and that no more than 10 $\%$ of 
the submillimetre population could be at z $>$ 4.  A few individual submillimetre galaxies 
have been found at z $>$ 4 (Wang et al 2007, Younger et al 2007, Schinnerer et al 2008), but there 
does not appear to be a large population of z $>$ 4 submillimetre galaxies.
The negative K-correction operating at submillimetre wavelengths
(Franceschini et al 1991, Blain and Longair 1993) means that dust-emitting galaxies of redshift 7
should have been clearly detectable in 850 $\mu$m surveys.  Presumably the low metallicity
at early times means star-forming regions are not opaque to dust so star-forming galaxies
at z = 5-7 emit purely in the optical and ultraviolet. 

The models are also consistent with the MAMBO 1200 $\mu$m integral counts of Greve et al (2004),
but integral counts do not strongly constrain model parameters because of the non-independence
of the errors.  {\it Herschel} and {\it Planck} submillimetre counts will provide much stronger tests of these models. 

\begin{figure*}
\epsfig{file=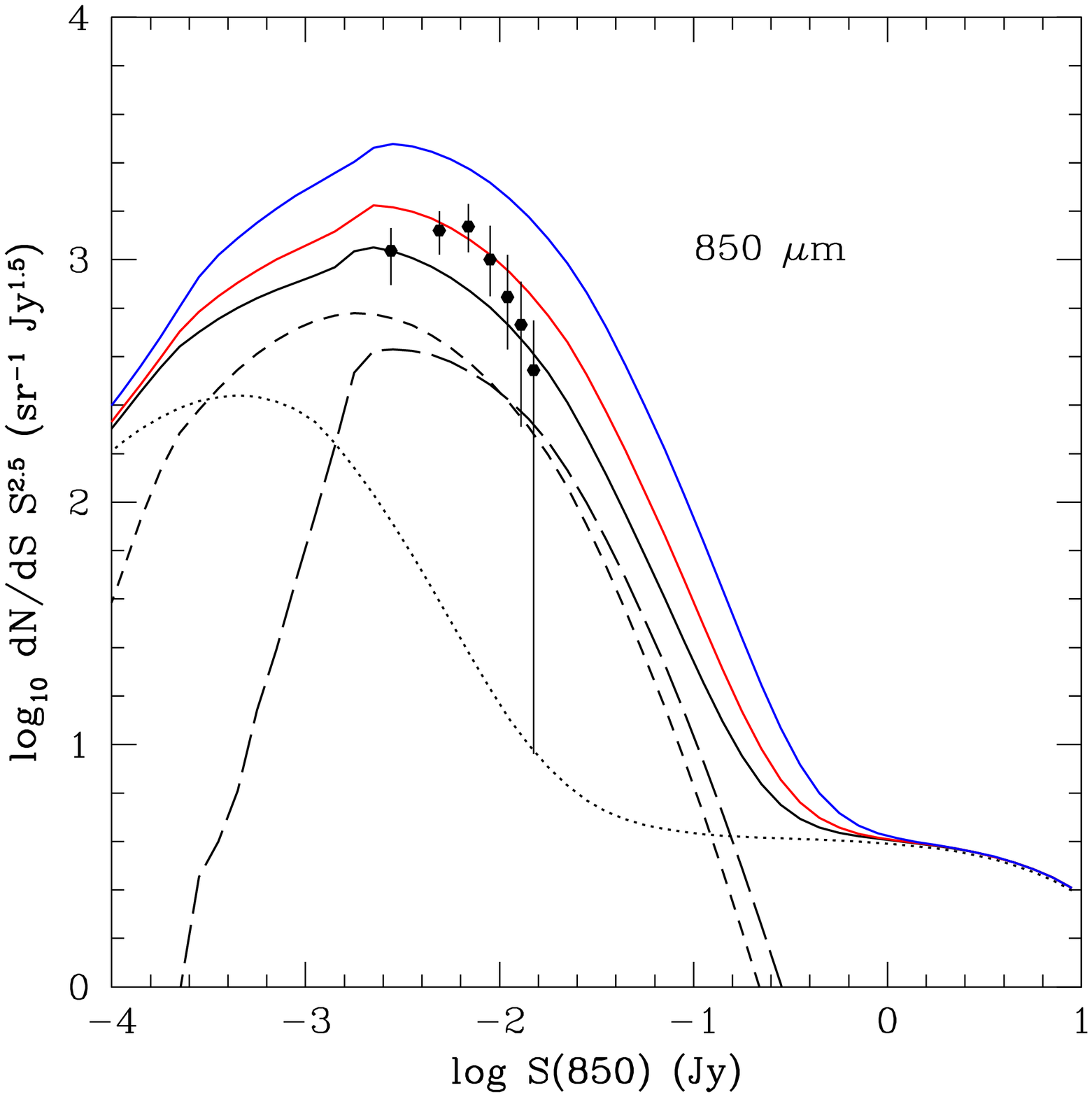,angle=0,width=7cm}
\epsfig{file=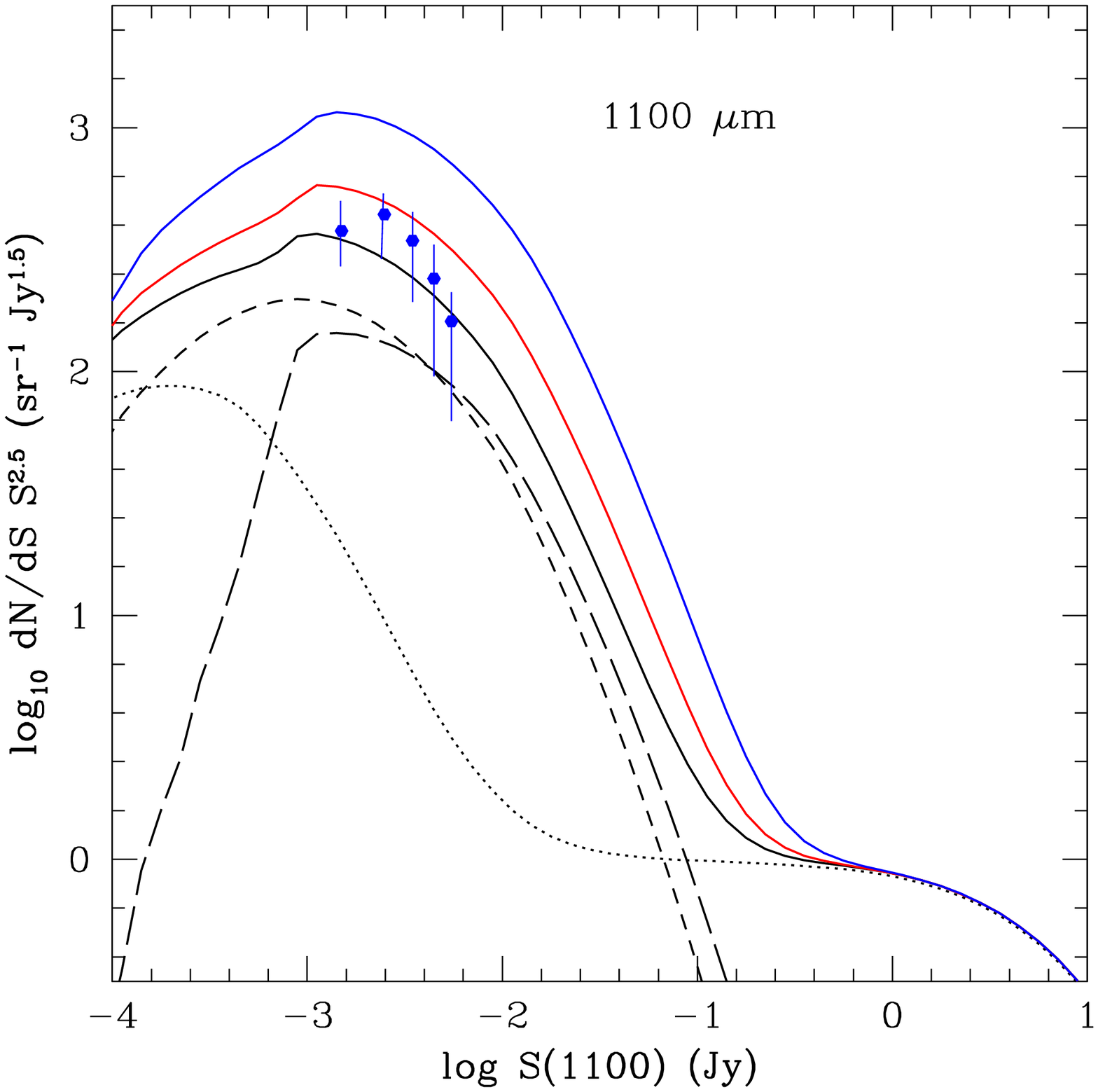,angle=0,width=7cm}
\caption{Euclidean normalised differential counts at 850 (LH) (SHADES data from  Coppin et al 2006)
and 1100 (RH) $\mu$m (
%black: SHADES-AZTEC data from Austermann et al 2008a, red: COSMOS data from
%Austermann et al 2008b, blue: 
GOODS data from Perera et al 2008).
Solid curve: total counts; dotted curve: cirrus; short-dashed curve: M82;
long-dashed curve: A220; dash-dotted curve: AGN dust tori.
Black curves: maximum redshift 4; red curve: maximum redshift 5; blue curve: maximum redshift 10.
}
\end{figure*}

Figure 7 shows a composite plot of differential counts at 8-1100 $\mu$m, each wavelength horizontally 
displaced for clarity.  Predictions for all these wavelengths (and others of interest for {\it Akari}, {\it Herschel} and
{\it Planck}) are available at
http:$//$astro.ic.ac.uk$/\sim$mrr$/$counts

\begin{figure*}
\epsfig{file=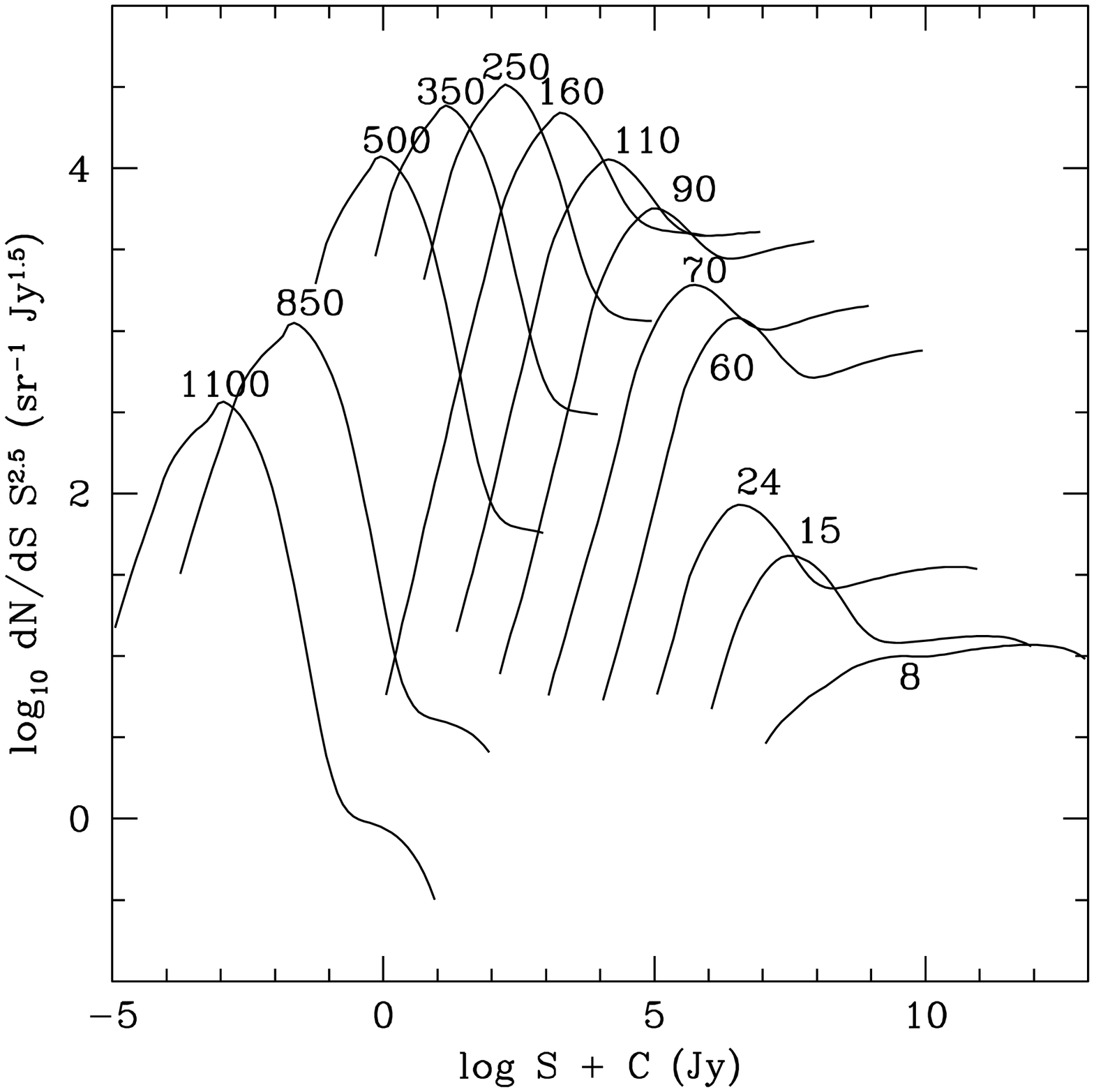,angle=0,width=14cm}
\caption{Euclidean normalised differential counts at wavelengths from 8-1100 $\mu$m, each 
displaced by 1 dex for clarity.
}
\end{figure*}

\section{Infrared background}

The infrared background spectrum provides a strong test of the assumed evolution and SEDs.
Figure 8 shows the fit of my best counts model to the infrared background spectrum, showing the effect
of assuming $z_f$ = 4, 5, 10.  The 100-850 $\mu$m background measurements require $z_f$ = 4-5,
with 4 giving the best fit.  So both the background spectrum and the 1100 $\mu$m counts are
both sensitive to $z_f$ and require $z_f$ = 4-5.

The fit to the background spectrum from 15-850 $\mu$m is excellent.  In Fig 8 I have shown the relative 
contribution of the different infrared poulations, with cirrus galaxies dominating at 8-24 $\mu$m, M82
starbursts dominating to 100 $\mu$m, and then cirrus galaxies again dominating in the submillimetre,
by virtue of their dominance of the very faint source counts.  The Arp220, high optical depth, starbursts,
which are the dominant contribution to the infrared luminosity function above $10^{12} L_{\odot}$, 
contribute only about 10$\%$ of the submillimetre background.  AGN dust tori contribute about 0.5 $\%$
of the mid infrared background and a negligible fraction of the submillimetre background.

\begin{figure*}
\epsfig{file=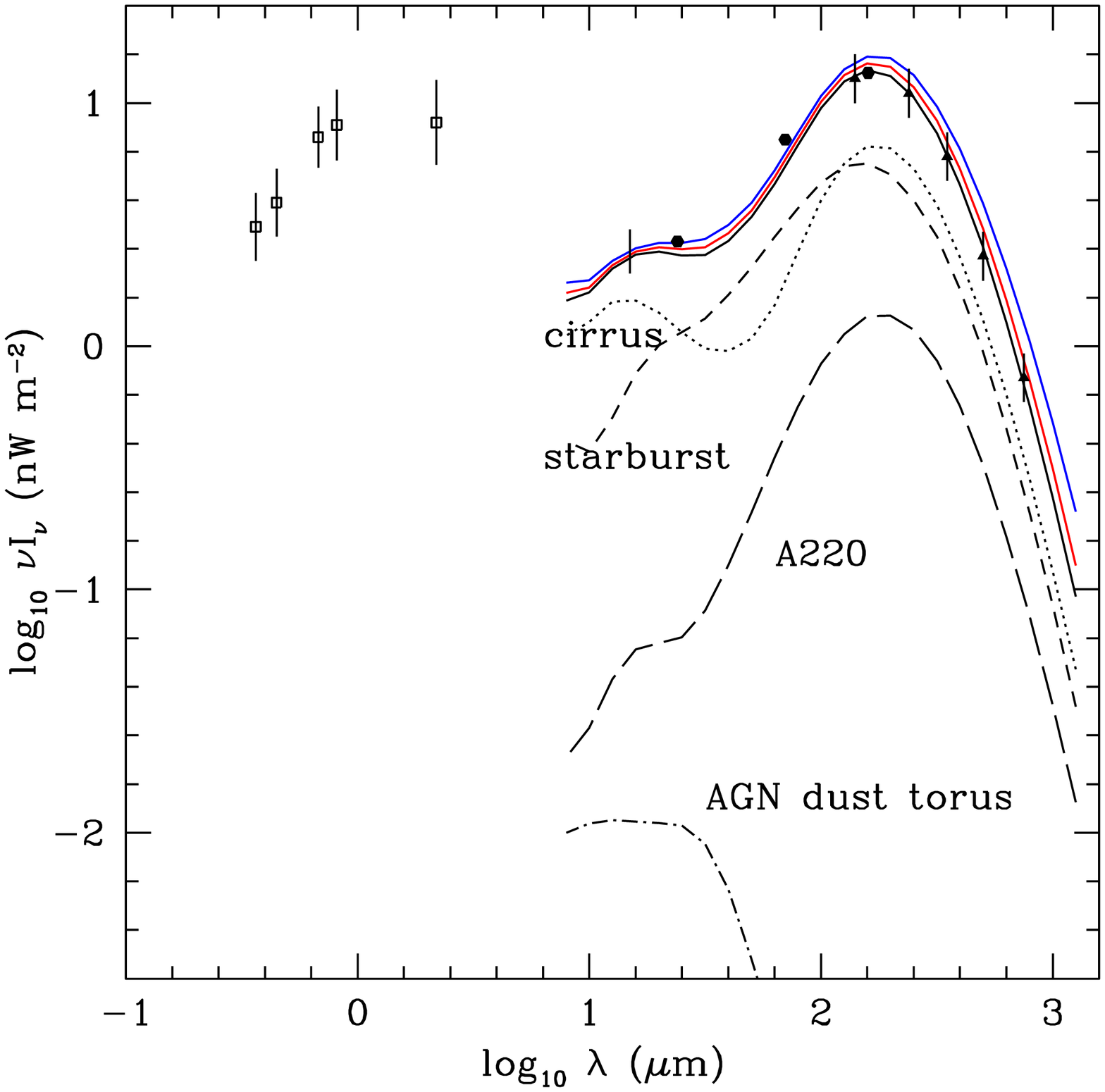,angle=0,width=14cm}
\caption{Predicted integrated background spectrum in infrared and submillimetre for maximum redshift
4 (black), 5 (red), 10(blue) showing contribution of different template populations.  Data from Fixsen et al (1998, open triangles),
Dole et al (2006, filled circles), Serjeant et al (2000, vertical bar), Pozzetti et al (1998, open squares).
}
\end{figure*}

\section{Predicted redshift distributions}

Detailed predictions of the models are given at a wide range of wavelenghs from 8-1100 $\mu$m,
including the survey wavelengths of Planck and Herschel (astro.ic.ac.uk$/\sim$mrr$/$counts$/$).  
These include the redshift distribution for each infrared type at each flux.

Figure 9 illustrates redshift distributions at 200 $\mu$Jy at 24 $\mu$m and at 5 mJy at  850 $\mu$m. 
The 850 $\mu$m distribution agrees rather well with the Chapman et al (2003, 2005) measured distribution.
The predicted 24 $\mu$m distribution shows far more z $>$ 1 sources than in the distribution derived
by Rowan-Robinson et al (2008) from their SWIRE photometric redshift catalogue, reflecting the
rather shallow optical photometry for most of that catalogue.

\begin{figure*}
\epsfig{file=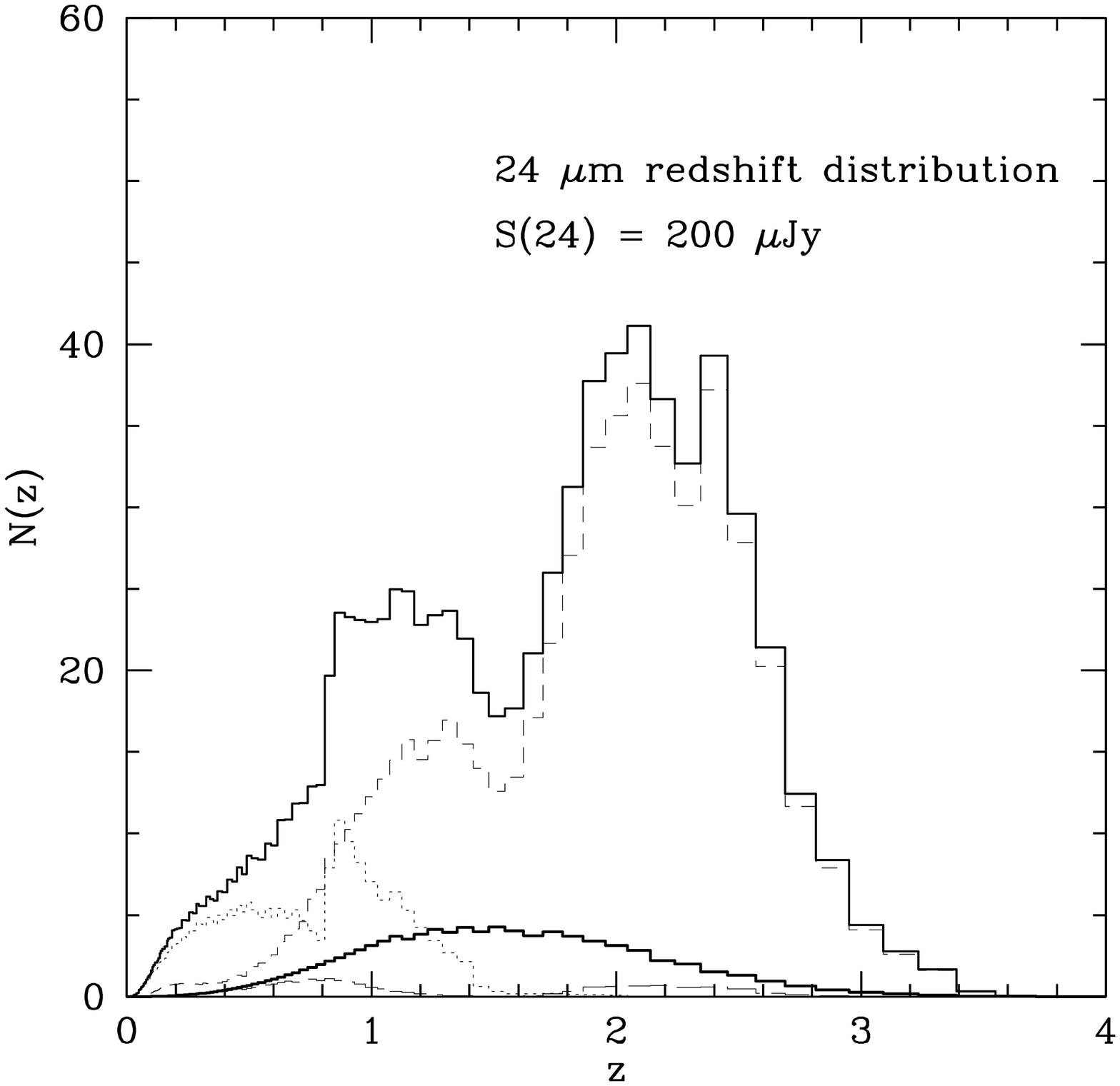,angle=0,width=7cm}
\epsfig{file=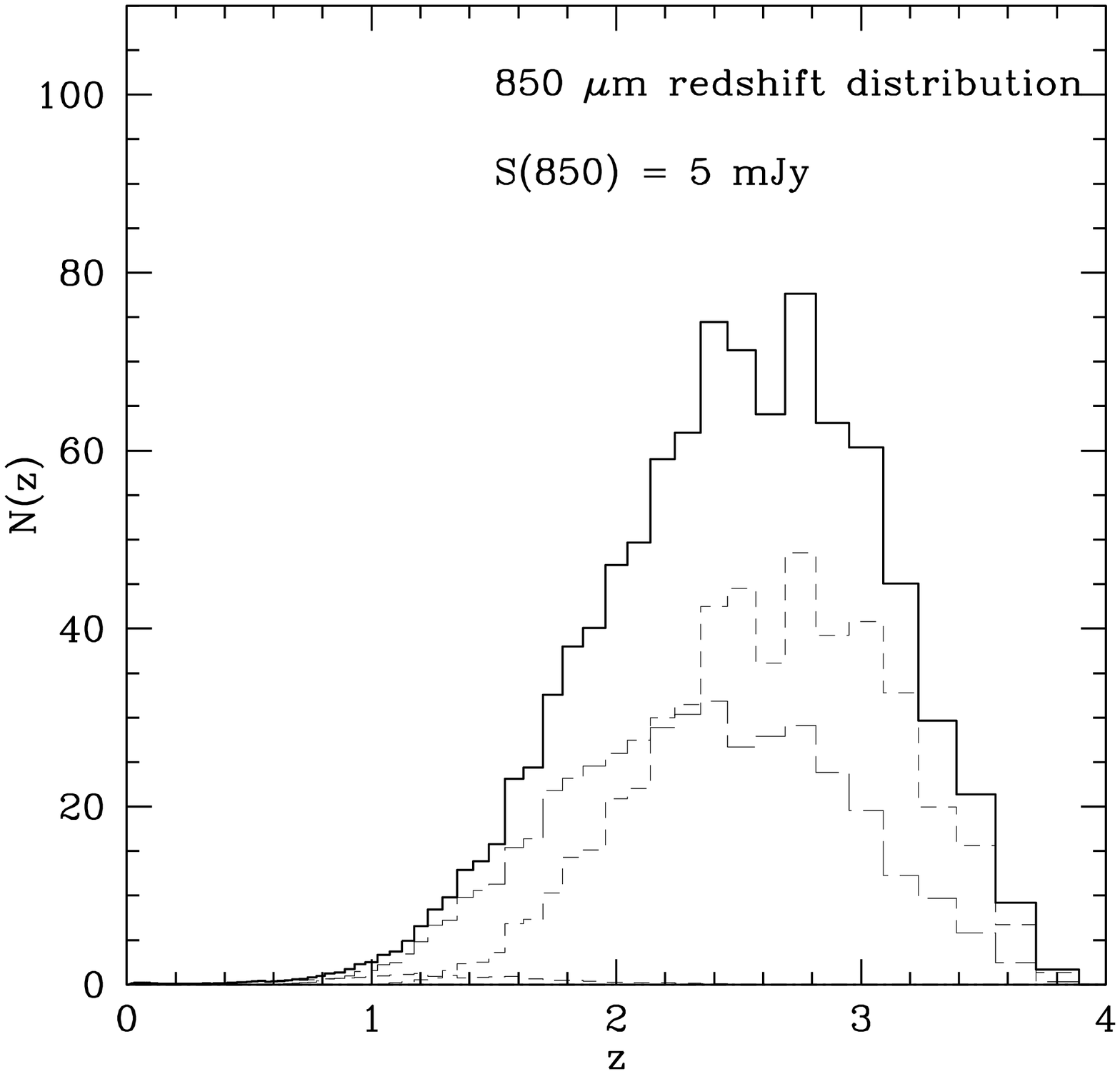,angle=0,width=7cm}
\caption{Redshift distribution at 24 $\mu$m (LH) and 850 $\mu$m (RH),
at S(24) = 200 $\mu$Jy and S(850) = 5 mJy, respectively.}
\end{figure*}

\section{Discussion}

The model presented here, based on four basic infrared templates is obviously an oversimplification.  
We know that to understand local galaxies we need a range of cirrus components (Rowan-Robinson 1992, 
Efstathiou and Rowan-Robinson 2003).  
From the models of Efstathiou et al (2000) the SED of a starburst varies strongly with the age of the starburst
so the representation as two simple extremes, M82 and Arp220 starbursts, is oversimplified.
Finally we know that we need a range of dust torus models to fit ultraluminous
and hyperluminous galaxies (Efstathiou and Rowan-Robinson 1995, Rowan-Robinson 2000, Farrah et al 2003). 

However the four templates used here do capture the main features of the infrared galaxy population.  In fitting
the SEDs of ISO and {\it Spitzer} sources (Rowan-Robinson et al 2004, 2005, 2008), we allow individual sources
to include cirrus, M82 or Arp220 starburst, and an AGN dust torus, which gives a rich range of predicted
SEDs.  This works for SCUBA sources (Clements et al 2008) and for detailed IRS spectroscopy of
infrared galaxies (Farrah et al 2008).

The strongest evolution rate is found for M82 starbursts.  Arp220 starburst peak at a slightly later redshift
(2 rather than 2.5) and at a slightly lower amplitude.   AGN dust tori, where the evolution reflects the 
accretion history, also peak at redshift 2, but show a less steep decline to the present than M82 or Arp220
starbursts, suggesting a gradual change in the relative efficiency with which gas is converted to stars
or accreted into a central black hole.  

Cirrus galaxies show a much more modest evolution rate, peaking at redshift 1, and with an amplitude
only a factor of 3 greater than the present epoch.  However even in relatively quiescent galaxies like our
own or M81, the star formation activity was greater at redshift 1 than at the present day.   This would also
imply a higher interstellar medium dust optical depth in spirals at z $\sim$ 1, an effect that may have been 
detected (Rowan-Robinson 2003, Rowan-Robinson et al 2008).

Lagache et al (2004) modified their 2003 model (Lagache et al 2003) for far infrared counts and the background 
spectrum with changes to the luminosity-density
evolution and to the 2-30 $\mu$m SED of their starburst component.  Their earlier model involves modest density evolution
for normal cirrus galaxies from z = 0 to 0.4, then constant luminosity-density to z = 5, followed by a slow decline to z  = 8.
For starbursts the luminosity-density rises steeply to z = 0.5, is relatively flat to z = 4, and then declines to redshift 8.
Lagache et al (2004) also provide successful fits to counts from 15-850 $\mu$m
and the background spectrum, with a philosophy rather different from the present paper.
The evolution function involves several discontinuous changes, a feature I have avoided in the present work.
I see no necessity to modify our well-tested SED models for 
starburst galaxies.  Abundant {\it Spitzer}-IRS data supports our assumed templates (eg Farrah et al 2008).

\section{Conclusions}

I have presented a new model for source counts from 8-1100 $\mu$m, which agrees well with source-count
data and the observed background spectrum.  The model is similar to that of Rowan-Robinson (2001), but
with different evolution for each of the four assumed infrared template types.   The evolution is modified in two
ways; (i) the exponential factor is modified so that it tends to a constant value at late times, (ii) the power-law
factor is modified so that it tends to zero at redshift $z_f$, rather than 0 as assumed by Rowan-Robinson (2001).
I find strong evidence from the 850 and 1100 $\mu$m counts and from the infrared background that $z_f$ = 4-5, with
some preference for a value at the low end of the range, implying
that star-forming galaxies at z $>$ 5 are not significant infrared emitters, presumably due to a low opacity in
dust at these early epochs.

The model involves zero or even negative evolution for starbursts and AGN at low redshifts ($<$0.2), suggesting
that the era of major mergers and strong galaxy-galaxy interactions is over.

{\it Herschel} and {\it Planck} submillimetre counts will provide much stronger tests of these models.

\section{Acknowledgements}

I thank the referee for helpful comments which helped me improve the paper. 

%\clearpage


\begin{thebibliography}{99}

%\bibitem{} Adelberger K.L., Steidel C.C., 2000, ApJ 544, 218

%\bibitem{} Afonso-Luis A. et al, 2004, MNRAS 354, 961 

%\bibitem{} Arnouts S., Christiani S., Moscardini L., Matarrese S., Lucchin F., Fontana A., Giallongo E., 
%1999, MN 310, 540

%\bibitem{} Aussel H., Cesarsky C.J., Elbaz D., Starck J.L., 1999, AA 342, 313

\bibitem{} Aussel H. et al, 1999, AA351, 37

\bibitem{} Austermann J.E., et al, 2008, MNRAS (to be submitted)

%\bibitem{} Babbedge T.S.R., Rowan-Robinson M., Gonzalez-Solares E., Polletta M., Berta S., Perez-Fournon I.,
%Oliver S., Salaman D.M., Irwin M., Weatherley S.J., 2004, MNRAS 353, 654

%\bibitem{} Babbedge T.S.R. et al,  2006, MNRAS 370, 1159

%\bibitem{} Ball N.M., Brunner R.J., Myers A.D., Strand N.E., Alberts S.L., Tcheng D., Llora, X., 2007, ApJ 663, 774

\bibitem{} Balland C., DevriendtJ.E.G., Silk J., 2003, MNRAS 343, 107

%\bibitem{} Barger A.J., Cowie L.L., Sanders D.B., Fulton E., Taniguchi Y., Sato Y., Kawara K., 
%Okuda H., 1998, Nat 394, 248

\bibitem{} Barger A.J., Cowie L.L., Sanders D.E., 1999, ApJ 518, L5

%\bibitem{} Barger A.J., Cowie L.L., Richards .A., 2001, ApJ (in press), astro-ph/0001096

%\bibitem{} Baugh C., Cole S., Frenk C., Lacey C., 1998, ApJ 498, 504

%\bibitem{} Bell E.F., 2003, ApJ 586, 794

%\bibitem{} Benitez N., 2000, ApJ 536, 571

%\bibitem{} Benitez N., et al, 2004, ApJS 150, 1

%\bibitem{} Berta S., Fritz J., Franceschini A., Bressan A., 2004, AA 418, 913

%\bibitem{} Berta S., Lonsdale C.J., Siana B., Farrah D., Smith H.E., Polletta M., Franceschini A., Fritz J., 
%Perez-Fournon I., Rowan-Robinson M., Shupe D., Surace J., 2007, AA 467, 565

%\bibitem{} Bertin E., Dennefeld M., Moshir M., 1997, AA 323, 685

%\bibitem{} Bolzonella M., Miralles J.-M., Pello R., 2000, AA 363, 476

\bibitem{} Borys C., Scott D., Chapman S.C., Halpern M., Nandra K., Pope A., 2004, MNRAS 355, 485

%\bibitem{} Brodwin M. et al, 2006, ApJ 651, 791

\bibitem{} Bruzual A.G., and Charlot, S., 1993, ApJ 405, 538

\bibitem{} Blain A.W., Longair M.S., 1993, MNRAS 264, 509

\bibitem{} Blain A.W., Smail I., Ivison R.J., Kneib J.-P., 1999, MNRAS 302, 632

%\bibitem{} Le Borgne D. and Rocca-Volmerange B., 2002, AA 386, 446

%\bibitem{} Buat V., Donas J., Deharveng J.-M., 1987, AA 185, 33

%\bibitem{} Budavari T., Szalay A.S., Connolly A.J., Csadai I., Dickinson M., 2000, AJ 120, 1588

%\bibitem{} Bruzual A.G., Charlot S., 1993, ApJ 405, 538

%\bibitem{} Burigana C., Danese L., de Zotti G., Franceschini A., Mazzei P., Toffalatti L., 1997, MNRAS 287, L17

%\bibitem{} Calzetti D., 1997, AJ 113, 162

%\bibitem{} Calzetti D., 1998, in 'Dwarf Galaxies and Cosmology', ed. T.X.Thuan, C.Balkowski, V.Cayatte,
%J.T Thanh Van (Editions Frontieres) asto-ph/9806083

%\bibitem{} Calzetti D., and Kinney A.L., 1992, ApJ 399, L39

%\bibitem{} Calzetti D., Heckman T.M., 1999, ApJ 519, 27

\bibitem{} Capak P. et al, 2008, ApJ 681, L53

\bibitem{} Chapman S.C., Blain A.W., Ivison R.J., Smail I.R., 2003, Nature 422, 695

\bibitem{} Chapman S.C., Blain A.W., Ivison R.J., Smail I.R., 2005, ApJ 622, 772

\bibitem{} Chary R. and Elbaz D., 2001 ApJ 556, 562

\bibitem{} Chary R., Casertano S., Dickenson M.E., Ferguson H.C., Eisenhardt P.R.M., Elabaz D., Grogin N.A., Moustakas L.A., Reach W.T., Yan H.,  2004 ApJS 154, 80

\bibitem{} Clements D. et al, 2008, MNRAS 387, 247

%\bibitem{} Cohen J.G., Hogg D.W., Blandford R., Cowie L.L., Hu E., Songaila A., Shopbell P., Richberg K., 
%2000, ApJ 538, 29

%\bibitem{} Cohen J.G., 2001, AJ 121, 2895

%\bibitem{} Cole S., Aragon-Salamanca A., Frenk C.S., Navarro J.F., Zepf S.E., 1994, MN 271, 781

%\bibitem{} Collister A.A. and Lahav O., 2004, PASP 116, 345

%\bibitem{} Connolly A.Z., Szalay A.S., Dickinson M., Subbarao M.U., Brunner P.J., 1997, ApJ 486, L11

\bibitem{} Coppin K. et al, 2006, MNRAS 372, 1621

\bibitem{} Cowie L.L., Barger A.J., Knieb J.-P., 2002, AJ 123, 2197

%\bibitem{} Cowie L.L., Songaila A., Barger A.J., 2000, AJ (in press), astro-ph/9904345

%\bibitem{} Cram L., 1998, ApJ 507,155

\bibitem{} Dannerbauer H., Walter F., Morrison G., 2008, ApJ 673, L127

%\bibitem{} Davoodi P. et al, 2006, MNRAS 371, 1113

%\bibitem{} Deharveng J.-M., Sasseen T.P., Buat V., Lampton M., Wu X., 1994, AA 289, 715

\bibitem{} Dole H., et al, 2001, AA 372, 364

\bibitem{} Dole H. et al, 2004, ApJS 154, 87

\bibitem{} Dole H., Lagache G., Puget J.-L., Caputi K.I., Fernandez-Conde N., Le Floc'h C.,
Papovich C., Perez-Gonzalez P.G., Rieke G.H., Blaylock M., 2006, AA 451, 417

\bibitem{} Dwek E., Arendt R.G., Hauser M.G., Fixsen D., Kelsall T., Leisawitz D., Pei Y.C., Wright E.L.,
Mather J.C., Moseley S.H., Odegard N., Shafer R., Silverberg R.F., Weiland J.L., 1998, ApJ 
508, 106

\bibitem{} Eales S., Lilly S.J., Gear W.K., Dunne L., Bond J.R., Hammer F., Le Fevre O., Crampton D., 
1999, ApJ 515, 518

\bibitem{} Eales S., Lilly S.J., Webb T., Dunne L., Gear W., Clements D., Yun M., 2000, AJ 120, 2244

\bibitem{} Efstathiou A., Rowan-Robinson M., 1995, MNRAS 273, 649

\bibitem{} Efstathiou A. et al, 2000a, MNRAS 319, 1169

\bibitem{} Efstathiou A., Rowan-Robinson M., Siebenmorgen R., 2000b, MNRAS 313, 734

\bibitem{} Efstathiou A., Rowan-Robinson M., 2003, MNRAS 343, 322

\bibitem{} Elbaz D. et al, 1999, AA 351, L37

\bibitem{} Elbaz D., Cesarsky C.J., Chanial P., Aussel H., Franceschini A., Fadda D., Chary R.R., 2002, 
AA 384, 848

%\bibitem{} Erb D.K., Shapley A.E., Steidel C.C., Pettini M., Adelberger K.L., Hunt M.P., Moorwood A.F.M.,
%Cuby J.-G., 2003, ApJ (in press), astro-ph/0303392

\bibitem{} Farrah D., Afonso J., Efstathiou A., Rowan-Robinson M., Fox M., Clements D., 2003, MNRAS 
343, 585

\bibitem{} Farrah D., Lonsdale C.J., Weedman D.W., Spoon H.W.W., Rowan-Robinson M., Polletta M.,
Oliver S., Houck J.R., Smith H.E., 2008, ApJ 677, 957

\bibitem{} Fazio G.G. et al, 2004, ApJ 154, 10

%\bibitem{} Fernandez-Soto A., Lanzetta K.M., Yahil A., 1999, ApJ 513, 34 

%\bibitem{} Fernandez-Soto A., Lanzetta K.M., Chen H.-W., Pascaelle S., Yahata N., 2001, ApJS 135, 41

%\bibitem{} Fernandez-Soto A., Lanzetta K.M., Chen H.-W., Levine B., Yahata N., 2002, MNRAS 330, 889

%\bibitem{} Firth A.E. et al, 2002, MNRAS 332, 617

%\bibitem{} Firth A.E., Lahav O., Sommerville R.S., 2003, MNRAS 339, 1195

\bibitem{} Fixsen D.J., Dwek E., Mather J.C., Bennett C.L., Shafer R.A., 1998, ApJ 508, 123

%\bibitem{} Flores H. et al, 1999, ApJ 517, 148

%\bibitem{} Fontana A., D'Odorico S., Poli F., Giallongo E., Arnouts S., Christiani S., Moorwood A.,
%Saracco P., 2000, MNRAS 339, 260

\bibitem{} Fox M.J. et al, 2002, MNRAS 331, 839

%\bibitem{} Franceschini A., Danese L., de Zotti G., Xu F., 1988, MNRAS 233, 175

\bibitem{} Franceschini A., Toffolatti L., Mazzei P., Danese L., De Zotti G., 1991, AAS 89, 285 

\bibitem{} Franceschini A.,  Mazzei, P.,  de Zotti, G.,  Danese, L., 1994, ApJ 427, 140

\bibitem{} Franceschini A., Silva L., Fasano G., Granato G.L., Bressan A., Arnouts S., Danese L., 
1998, ApJ  506, 600

\bibitem{} Franceschini A., Aussel H., Cesarsky C.J., Elbaz D., Fadda D., 2001, AA 378, 1

%\bibitem{} Franceschini A. et al, 2007, (S1)

\bibitem{} Frayer D.T., Huynh M.T., Chary R., Dickinson M., Elbaz D., Fadda D., Surace J.A., 
Teplitz H.I., Yan L., Mobasher B., 2006, ApJ 647, L9

%\bibitem{} Furusawa H. et al, 2008, ApJS (in press), astro-ph/0801.4017

%\bibitem{} Gabasch A. et al, 2004, AA 421, 41

%\bibitem{} Gavignaud I. et al, 2006, AA 457, 79

%\bibitem{} Gallego J., Zamorano J., Aragon-Salamanca A., Rego M., 1995, ApJ 455, L1

\bibitem{} Gispert R., Lagache G., Puget J.-L., 2000, AA 360, 1

%\bibitem{} Glazebrook K., Blake C., Economou F., Lilly S., Colless M., 1999, MN 306, 843

%\bibitem{} Glazebrook K. et al, 2001

%\bibitem{} Gregg M.D., Lacy M., White R.L., Glikman E., Helfand D., Becker R.H., Brotherton M.S., 2002, ApJ 564, 133

%\bibitem{} Gregorich D.T., Neugebauer G., Soifer B.T., Gunn J.E., Herter T.L., 1995, AJ 110, 259

\bibitem{} Greve T.R., Ivison R.J., Bertoldi F., Stevens J.A., Dunlop J.S., Lutz D., Carilli C.L., 2004, MNRAS 354, 779

%\bibitem{} Gronwall C., 1998, in 'Dwarf Galaxies and Cosmology', eds T.X.Thuan, C.balkowski, 
%V.Cayatte, J.Tran Thanh Van, (Editions Frontieres), astro-ph/9806240

\bibitem{} Gruppioni C., Lari C., Pozzi F., Zamorani G., Franceschini A., Oliver S., Rowan-Robinson M., Serjeant S.,
2002, MNRAS 335, 831

\bibitem{} Guiderdoni B., Bouchet F.R., Puget J.-L., Lagache G., Hivon E., 1997, Nature 390, 257

\bibitem{} Guiderdoni B, Hivon E., Bouchet F.R., Maffei B., 1998, MNRAS 295, 877

%\bibitem{} Gwynn S.D.J. and Hartwick F.D.A., 1996, ApJ 468, L77

%\bibitem{} Haarsma D.B., Partridge R.B., Windhorst R.A., Richards E.A., 2000, ApJ 544, 641

\bibitem{} Hacking P.B. and Houck J., 1987, ApJS 63, 311

%\bibitem{} Hammer F., Flores H., 1998, 
%in 'Dwarf Galaxies and Cosmology', eds T.X.Thuan, C.Balkowski, V.Cayatte, J.Tran Thanh Van, 
%(Editions Frontieres), astro-ph/9806184

\bibitem{} Hauser M.G., Arendt R.G., Kelsall T., Dwek E., Odegard N., Weiland J.L., Freudenreich H.T.,
Reach W.T., Silverberg R.F., Moseley S.H., Pei Y.C., Lubin P., Mather J.C., Shafer R.A., Smoot G.F.,
Weiss R., Wilkinson D.T., Wright E.L., 1998, ApJ 508, 25

\bibitem{} Heradeau Ph., et al, 2004, MNRAS 354, 924

%\bibitem{} Hogg D.W., et al, 1998, AJ 115, 1418

%\bibitem{} Hopkins et al, 2001, AJ 122, 288

%\bibitem{} Hsieh B.C., Lee H.K.C., Lin H., Gladders M.D., 2005, ApJS 158, 161

\bibitem{} Hughes D.H., Serjeant S., Dunlop J., Rowan-Robinson M., Blain A., Mann R.G., Ivison R.,
Peacock J., Efstathiou A., Gear W., Oliver S., Lawrence A., Longair M., Goldschmidt P., Jenness T.,
1998, Nat 394, 241

\bibitem{} Ibar E., Cirasuolo M., Ivison R., Best P., Smail I., Biggs A., Simpson C., Dunlop J., Almaini O., McLure R.,
Foucaud S., Rawlings S., 2008 MNRAS 386, 953

%\bibitem{} Ilbert O. et al, 2006, AA 457, 8411

%\bibitem{} Irwin M., Lewis J., 2001, New Astronomy Reviews 45, 1051

\bibitem{} Ivison R.J., et al, 2002, MN 337, 1

\bibitem{} Kawara K., Sato Y., Matsuhara H., Taniguchi Y., Okuda H., Sofue Y., Matsumoto T.,
Wakamatsu K., Karoji H., Okamura S., Chambers K.C., Cowie L.L., Joseph R.D., and Sanders D.B., 1998, 
AA 336, L9

\bibitem{} King A.J., Rowan-Robinson M., 2003, MNRAS 339, 260

%\bibitem{} Kinney A.L., Bohlin R.C., Calzetti D., Panagia N., Wyse R.F.G., 1993, ApJS 86, 5

%\bibitem{} Lacy M. et al, 2004, ApJS 154, 166

%\bibitem{} Lanzetta K.M., Yahil A., Fernandez-Soto A., 1996, Nature 381, 759

%\bibitem{} Lanzetta K.M., Yahata N., Pascarelle S., Chen H.-W., Fernandez-Soto A., 2002, ApJ 570, 492

%\bibitem{} Leech K.J., Penston M.V., Terlevich R., Lawrence A., Rowan-Robinson M., Crawford J., 
%1989, MN 240, 349

%\bibitem{} Le F\'evre O. et al, 2004, AA 417, 839

%\bibitem{} Le F\'evre O. et al, 2003, The Messenger 111, 18

%\bibitem{} Le F\'evre O. et al, 2005, AA 439, 845

\bibitem{} Lagache G., Abergel A., Boulanger F., Desert F.-X., Puget J.-L., 1999, AA 344, 322

\bibitem{} Lagache G., Dole H., Puget J.-L., 2003, MNRAS 338, 555

\bibitem{} Lagache G., et al, 2004, ApJS 154, 112 

\bibitem{} Le Floch E. et al, 2005, ApJ 632, 169

\bibitem{} Lilly, S.J., Le Fevre, O., Hammer, F., Crampton, D., 1996, ApJ 460, L1

%\bibitem{} Liu et al 1998

\bibitem{} Lonsdale C.J., Hacking P.B., Conrow T.B., Rowan-Robinson M., 1990, ApJ 358, 60

%\bibitem{} Lonsdale C. et al, 2003, PASP 115, 897

%\bibitem{} Lonsdale C.J. et al, 2004, ApJS 154, 54

%\bibitem{} Madau, P., Ferguson, H.C., Dickinson, M.E., Giavalisco, M., Steidel, C.C., Fruchter, A., 
%1996, MNRAS 283, 1388

%\bibitem{} Madau, P., 1997, in 'Star Formation Near and Far', (Un.iv. of Maryland) astro-ph/9612157

%\bibitem{} Madau, P., 1998, in 'The Hubble Deep Field', ed. M.Livio, S.M.Fall and P.Madau (STScI 
%Symposium Series) astro-ph/9709147

%\bibitem{} Madau P., Pozzetti L., Dickinson M., 1998, ApJ 498, 106

%\bibitem{} Mann R.G. et al, 2001, MN (submitted)

%\bibitem{} Maraston C., 2005, MNRAS 362, 799

%\bibitem{} Maraston C., 2006, ApJ 652, 85

\bibitem{} Marleau F.R.l, 2004, ApJS 154, 66

%\bibitem{} Mazzarella J.M., Balzano V.A., 1986, ApJS 62, 751

%\bibitem{} McCracken H.J. et al, 2003, AA 410 17

%\bibitem{} McMahon R.G., Walton N.A., Irwin M.J., Lewis J.R., Bunclark P.S., Jones D.H.P., Sharp R.G., 2001,
%New Astronomy Reviews, astro-ph/0001285

\bibitem{} Menendez-Delmestre K., Blain A.W,m Alexander D.M., Smail I., Armus L., Chapman S.C., Frayer D.T.,
Ivison R.,J., Teplitz H.I., 2007, ApJ 655, L65

%\bibitem{} Meurer G.R., Heckman T.M., Leitherer C., Kinney A., Robert C., Garnett D.R., 1995, 
%AJ 110, 2665

%\bibitem{} Meurer G.R., Heckman T.M., Lehnert M.D., Leitherer C., Lowenthal J., 1997, AJ 114, 54

%\bibitem{} Meurer G.R., 1998, in 'Hubble Deep Field', ed.  M.Livio, S.M.Fall and P.Madau (STScI 
%Symposium Series) astro-ph/9708163

%\bibitem{} Meurer G.R., Heckman T.M., Calzetti D., 1999, ApJ 521, 64

%\bibitem{} Mobasher, B., Rowan-Robinson, M., Georgakakis, A., Eaton, N., 1996, MNRAS 282, L7

%\bibitem{} Mobasher M. and Mazzei P., 1999, in 'Photometric Redshifts and High Redshift Galaxies',
%Weymann et al. Eds., Astr.Soc. of Pac.Conf. Series 191, astro-ph/9907210

%\bibitem{} Mobasher B. et al, 2004, ApJ 600, L167

%\bibitem{} Mobasher B. et al, 2007, ApJS 172, 117

%\bibitem{} Oliver, S., {\it et al.}, 1995, in 'Wide-Field Spectroscopy and the Distant Universe', eds.
%S.J.Maddox and A.Arogon-Salamanca (World Scientific) p.274

\bibitem{} Oliver S. et al, 1997, MNRAS 289, 471

\bibitem{} Papovich C. et al, 2004, ApJS 154, 70

%\bibitem{} Pascarelle S.M., Lanzetta K.M., Fernandez-Soto A., 1998, ApJ 508, L1

%\bibitem{} Peacock J.A., Rowan-Robinson M., Blain A.W., Dunlop J.S., Efstathiou A., Hughes D.H.,
%Jenness T., Ivison R.J., Lawrence A., Longair M.S., Mann R.G., Oliver S.J., Serjeant S., 2000, MN 318, 535

\bibitem{} Pearson  C., Rowan-Robinson M., 1996, MNRAS 283, 174

%\bibitem{} Pei Y.C., Fall M.S, Hauser M.G., 1999, ApJ 522, 604

\bibitem{} Perera T.A., Chapin E.L., Austermann J.E., Scott K.S., Wilson G.W, Halpern M.,
Pope A., Scott D., Yun M.S., Lowenthal J.D., Morrison G. et al, 2008, MNRAS (in press), astro-ph$/$0806.3791

%\bibitem{} Pettini M., Steidel C.C., Adelberger K.L., Kellogg M., Dickinson M., Giavalisco M.,
%1997, in 'Cosmic Origins: Evolutionof Galaxies, Stars, Planets and Life, eds Woodward C.E., 
%Thronson H., Shull J.M., Astron.Soc.Pac.,
%(ASP Conf.Ser., San Francisco) in press (astro-ph/9708117)

%\bibitem{} Pettini M., Kellogg M., Steidel C.C., Dickinson M., Adelberger K.L.,, Giavalisco M., 
%1998, ApJ 508, 539

%\bibitem{} Pettini M., Steidel C.C., Kellogg M., Dickinson M., Adelberger KL., Giavalisco M., 1998, 
%in 'Dwarf Galaxies and Cosmology', eds T.X.Thuan, C.balkowski, V.Cayatte, J.Tran Thanh Van, 
%(Editions Frontieres)

%\bibitem{} Pei, Y.C., and Fall, S.M., 1995, ApJ 454, 69

%\bibitem{} Pei Y.C., Fall M., and Hauser M.G., 1999, ApJ 522, 604

%\bibitem{} Pierre M. et al, 2007, MNRAS 382, 279

%\bibitem{} Poggianti B.M., Bressan A., Franceschini A., 2001, ApJ 550, 195

%\bibitem{} Polletta M. et al, 2006, ApJ 642, 673

%\bibitem{} Polletta M. et al, 2007, ApJ 663, 81

\bibitem{} Pope A. et al, 2008, ApJ(in press), astro-ph/0808.2816

\bibitem{} Pozzetti L., Madau P., Ferguson H.C., Zamorani G., Bruzual G.A., 1998, MNRAS 298, 1133

\bibitem{} Puget, J.-L., Abergal A., Bernard J.-P., Boulanger F., Burton W.B., 
Desert F.-X., Hartmann D., 1996, AA 308, 5

%\bibitem{} Richards E.A. et al, 1998, AJ 116, 1039

%\bibitem{} Riess A.G. et al, 2001, ApJ (submitted), astro-ph/0104455

%\bibitem{} Rodighero G., Franceschini A., Fasano G., 2001, MN 324, 491

%\bibitem{} Rowan-Robinson M. et al, 1997, MN 289, 490

%\bibitem{} Reddy N.A., Steidel C.C., Fadda D., Yan Y., Pettini M., Shapley A.E., Erb D.K., Adelberger K.L., 2006, ApJ 653, 1004

%\bibitem{} Richards G.T., Weinstein M.A., Schneider D.P., Fan X., Strauss M.A., Vanden Berk D.E., 2001, AJ 122, 1151

%\bibitem{} Richards G.T. et al, 2003, AJ 126, 1131

\bibitem{} Rowan-Robinson M., Lawrence A., Saunders W., 1991, MNRAS 253, 485

\bibitem{} Rowan-Robinson M., 1992, MNRAS 258, 787

\bibitem{} Rowan-Robinson M., 1995, MNRAS 272, 737

\bibitem{} Rowan-Robinson M. et al, 1997, MNRAS 289, 490

\bibitem{} Rowan-Robinson M., 2000, MNRAS 316, 885

\bibitem{} Rowan-Robinson M., 2001, ApJ 549, 745

\bibitem{} Rowan-Robinson M., 2003, MNRAS 345, 819

%\bibitem{} Rowan-Robinson M., 2003b, MNRAS 344, 13

\bibitem{} Rowan-Robinson M. et al, 2004, MNRAS 351, 1290
 129, 1183

\bibitem{} Rowan-Robinson M. et al, 2005, AJ 129, 1183

\bibitem{} Rowan-Robinson M. et al, 2008, MNRAS 386, 697

%\bibitem{} Saunders, W., Rowan-Robinson, M., Lawrence, A., Efstathiou, G., Kaiser, N.,
%Frenk, C.S., 1990, MN 242, 318

%\bibitem{} Sawicki M.J., Lin H., Yee H.K.C., 1997, AJ 113, 1

\bibitem{} Schinnerer E., Carilli C.L., Capak P., Martinez-Sansigre A., Scoville N.Z., Smolcic V., Taniguchi Y.,
Yun M.S., Bertoldi F., Le Fevre O., de Ravel L., 2008, ApJL (in press), astro-ph/0810.3405

\bibitem{} Scott S.E. et al, 2002, MNRAS 331, 817

\bibitem{} Serjeant S. et al, 2000, MNRAS 316, 768

%\bibitem{} Sekiguchi et al, 2005 (http//www.springerlink.com/content/p4166v6530p86730/fulltext.pdf)

%\bibitem{} Siebenmorgen R., Krugel E., 2007, AA 461, 445

%\bibitem{} Shapley A.E., Steidel C.C., Adelberger K.L., Dickinson M., Giavalisco M., Pettini M., 2001,
%ApJ 562, 95

\bibitem{} Shupe D.L. et al, 2008, AJ 135, 1050

\bibitem{} Smail I., Ivison R.J., Blain A.W., 1997, ApJL 490, L5

\bibitem{} Smail I., Ivison R.J., Blain A.W., Kneib J.-P., 2002, MNRAS 331, 495

%\bibitem{} Smith E. et al, 2007 (in prep)

%\bibitem{} Steidel C.C., Adelberger K.L., Dickinson M., Giavalisco M., Pettini M., Kellogg M., 1998, ApJ 492, 428

%\bibitem{} Steidel C.C., Adelberger K.L., Giavalisco M., Dickinson M., Pettini M., 1999, ApJ 519, 1

%\bibitem{} SubbaRao M.U., Connolly A.J., Szalay A.S., 1996, AJ 112, 129

%\bibitem{} Sullivan M., Treyer M.A., Ellis R.S., Bridged T.J., Milliard B., Donas J., 2000,
%MN 312, 442

%\bibitem{} Sullivan M., Mobasher B., Chan B., Cram L., Ellis R., Treyer M., 
%Hopkins A., 2001, ApJ 558, 72

%\bibitem{} Steffen A.T., Barger A.J., Capak, P., Cowie L.L., Mushotzky R.F., Yang Y., 2004, AJ 128, 1483

%\bibitem{} Surace J. et al, 2004, The SWIRE N1 Image Atlases and Source Catalogs (Pasadena: Spitzer Science Center)

%\bibitem{} Swinbank A.M. et al, 2007, MNRAS 379, 1343

%\bibitem{} Telfer R.C., Zheng W., Gerard A., Davidsen A., 2002, ApJ 565, 773

%\bibitem{} Teplitz H.I., Hill R.S., Malumuth E.M., Collins N.R., Gardner J.P.,
%Palunas P., Woodgate B.E., 2001, ApJ 548, 127

%\bibitem{} Trammell G.B., Vanden Berk D.E., Schneider D.P., Richards G.T., Hall P.B., Anderson S.F., Brinkmann J., 2007, AJ 133, 1781 

%\bibitem{} Trichas M. et al, 2007 (in prep)

%\bibitem{} Thompson R.I., Weymann R.J., Storrie-Lombardi L.J., 2001, ApJ 546, 694

%\bibitem{} Tresse L., Maddox S.J., 1998, ApJ 495, 691 

%\bibitem{} Treyer M.A., Ellis R.S., Milliard B., Donas J., Bridges T.J., 1998, MN 300, 303

%\bibitem{} Vanzella E. et al, 2004, AA 423, 761

%\bibitem{} VanDen Berk D.E., 2001, AJ 122, 549

%\bibitem{}  van der Wel A., Franx M., van Dokkum P.G., HuangJ., Rix H.-W., Illingworth G.D., 2005, ApJ 636, L21

%\bibitem{} Vijh U.P., Witt A.N., Gordon K.D., 2003, astro-ph/0301121

\bibitem{} Wang W.H. et al, 2007, ApJ 670, L89

\bibitem{} Webb T.M. et al, 2003, ApJ 587, 41

%\bibitem{} Williams R.E. et al, 1996, AJ 112, 1335

%\bibitem{} Wilner D.O., Wright M.C.W., 1997, ApJ 488, L67

%\bibitem{} Wolf C., Meisenheimer K., Rix H.-W., Borch A., Dye S., Kleinheinrich M., 2003, AA 401, 73

%\bibitem{} Wolf C., Meisenheimer K., Kleinheinrich M., Rix H.-W., Borch A., Dye S., 2004, AA 421, 913

%\bibitem{} Xu C., Hacking P.B., Fang F., Shupe D.L., Lonsdale C.J., Lu N.Y., Hekou G., Stacey G.J.,
%Ashby M.L.N., 1998, ApJ 508, 576

%\bibitem{} Xu C., 2000, ApJ 541, 134

\bibitem{} Xu C., Lonsdale C.J., Shupe D.L., Franceschini A., Martin C., Schiminovich D., 2003, ApJ 587, 90

%\bibitem{} Yan L., McCarthy P.J., Freudling W., Teplitz H.I., Malumuth E.M., Weymann R.J., Malkan M.A.,
%1999, ApJ 519, L47

\bibitem{} Yoshii Y. and Takahara F., 1988, ApJ 326, 1

\bibitem{} Younger J.D. et al, 2007, ApJ 671, 1531

\end{thebibliography}
\end{document}